\documentclass[Crown,times]{sagej}

\usepackage[hyphens]{url}
\usepackage{moreverb,url}

\usepackage[colorlinks,bookmarksopen,bookmarksnumbered,citecolor=red,urlcolor=red]{hyperref}

\newcommand\BibTeX{{\rmfamily B\kern-.05em \textsc{i\kern-.025em b}\kern-.08em
		T\kern-.1667em\lower.7ex\hbox{E}\kern-.125emX}}

\def\volumeyear{2017}

\usepackage{pgfplots}
\usepackage{tikz}
\usetikzlibrary{patterns}

\usepackage{smartdiagram}
\smartdiagramset{set color list={
		gray!30,
		gray!30,
		gray!30,
		gray!30,
		gray!30,
		gray!30,
	},
	sequence item border color=black,
	sequence item font size=16,
}

\usepackage{float}

\usepackage{color}
\usepackage{colortbl}

\usepackage{amssymb}
\usepackage{comment}
\usepackage[multiple]{footmisc}
\usepackage{amsmath}

\usepackage{lipsum}
\usepackage{pfnote}
\usepackage{multirow}

\usepackage{pgfplots}
\usepackage{tikz}
\usetikzlibrary{arrows,decorations.pathmorphing,fit,positioning}
\usepackage{xcolor}
\usetikzlibrary{patterns}
\usepackage{amstext}

\usepackage{fixfoot} 
\DeclareFixedFootnote{\myfnone}{one}  

\usepackage{enumitem}

\usepackage{colortbl}

\usepackage{bbding} 

\definecolor{maroon}{cmyk}{0,0.87,0.68,0.32}

\usetikzlibrary{positioning,calc}

\usepackage{tikz}
\usetikzlibrary{trees}

\tikzset{level 1/.style={level distance=2cm, sibling distance=10cm}}
\tikzset{level 2/.style={level distance=2cm, sibling distance=3cm}}

\tikzset{bag/.style={text width=10em, text centered,yshift=-0.5cm}}

\usepackage{smartdiagram}
\usesmartdiagramlibrary{additions}

\usetikzlibrary{positioning,calc}

\def\addlegendimage{\csname pgfplots@addlegendimage\endcsname}

\begin{document}
	
	\runninghead{Karami et al.}
	
	\title{Twitter Speaks: A Case of National Disaster Situational Awareness}
	
	\author{Amir Karami\affilnum{1}, Vishal Shah\affilnum{2}, Reza Vaezi\affilnum{3}, Amit Bansal\affilnum{2}}
	
	\affiliation{\affilnum{1}School of Library and Information Science, University of South Carolina, USA \\ \affilnum{2}Computer Science \& Engineering Department, University of South Carolina, USA \\
		\affilnum{3}Coles College of Business, Kennesaw State University, USA}
	
	\corrauth{Amir Karami}
	\email{karami@sc.edu}
	
	\begin{abstract}
		In recent years, we have been faced with a series of natural disasters causing a tremendous amount of financial, environmental, and human losses. The unpredictable nature of natural disasters’ behavior makes it hard to have a comprehensive situational awareness (SA) to support disaster management. Using opinion surveys is a traditional approach to analyze public concerns during natural disasters; however, this approach is limited, expensive, and time-consuming. Luckily the advent of social media has provided scholars with an alternative means of analyzing public concerns. Social media enable users (people) to freely communicate their opinions and disperse information regarding current events including natural disasters. This research emphasizes  the value of social media analysis and proposes an analytical framework: Twitter Situational Awareness (TwiSA). This framework uses text mining methods including sentiment analysis and topic modeling to create a better SA for disaster preparedness, response, and recovery. TwiSA has also effectively deployed on a large number of tweets and tracks the negative concerns of people during the 2015 South Carolina flood.
	\end{abstract}
	
	
	\maketitle
	
	\section{Introduction}
	\label{Int}

	The growth of social media has provided an opportunity to track people's concerns in a new way. Twitter has 316 million users \cite{techrunch}, provides capability of real-time feedback, and utilizes time stamp to provide conversation updates to users. Therefore, Twitter's potential as a reliable and relevant as a data source is evident and provides a unique opportunity to understand users' concerns \cite{mejova2015twitter,paul2011you}. In contrast to surveys and surveillance networks that can take weeks or even years to collect data, Twitter is publicly accessible with no waiting time.

	SA is ``all knowledge that is accessible and can be integrated into a coherent picture, when required, to assess and cope with a situation" \cite{sarter1991situation}. Obtaining reliable and accurate information and providing access to knowledge derived from the collected data is the main concern in situational awareness. Thus, SA and its inherent processes play an essential role in helping people facing a natural disaster. During a natural disaster, people immediately try to collect or share information to keep them away from possible dangerous situations \cite{castillo2016big}. Developing high level of Situational Awareness (SA) by analyzing real-time big data helps to become more effective in disaster management \cite{kitchin2014real,madey2007enhanced}.

	In October 2015 hurricane Joaquin led to an unusually high amount of rainfall in the state of South Carolina (SC). The precipitation ranged from 11 inches to 26 inches in different areas and caused server flooding in the state. This flood caused more than \$12 billion damage to roads, homes, and infrastructures with 19 dead \cite{zurich,scemd}. Twitter users expressed their feelings and opinions during the 2015 SC flood through their Tweets. This valuable time-sensitive data can improve SA development and give all involved parties a more accurate picture of the situation during the flood. 
	
	Crowdsourced data such as social media data is a rapid and accessible source for SA development \cite{vieweg2010microblogging}. In other words, understanding users' concerns and their emotions can help disaster managers develop a better real-time SA and decision making plans \cite{neppalli2017sentiment,callaghan2016disaster}. During the height of the 2015 SC flood, thousands of people were twitting about the flood on an average of 3000 tweets per hour. One can imagine that manually analyzing this many tweets is not easily possible; therefore, new customized approaches are necessary to disclose hidden semantic features from social media. This study proposes an analytical framework, \textit{\textbf{\underline{Twi}}tter \textbf{\underline{S}}ituational \textbf{\underline{A}}wareness (TwiSA)}, to analyze unstructured flood-related tweets to better understand  temporal concerns of people affected by this natural disaster. The proposed framework employs both sentiment analysis and topic modeling to uncover temporal patterns of public concerns to provide a better SA. We apply TwiSA on a large number of tweets and tracks the negative concerns of people during the 2015 SC flood. 
	
	\section{Related Work}
	\label{RW}

	Twitter data has been used for a broad range of applications such as business \cite{bollen2011twitter,collins2018social,karami2018us,kordzadeh2019exploring,kordzadeh2018investigating}, politics \cite{tumasjan2010predicting,karami2018mining, alperin2018politicians, kitzie2018life, najafabadi2018hacktivism,karami2019political}, and health \cite{szomszor2010swineflu,webb2018Characterizing, karami2018characterizing,karami2018characterizingtrans, shaw2017computational,shaw2019exploratory}. In this section, we review the applications of Twitter data in natural disasters including fire, flood, earthquake, hurricane and typhoon, and volcano eruption.

	Abel et al. (2012) developed a tool, called Twitcident, to collect fire-related tweets and search tweets using queries \cite{abel2012twitcident}. Sinnappan et al. (2010) collected tweets and used qualitative approach to find meaningful categorization of fire-related tweets \cite{sinnappan2010priceless}
	
	Earthquake-related studies proposed frameworks to improve disaster management process. These included using data mining and natural language processing for damage detection and assessment of earthquakes \cite{avvenuti2014ears}, proposing a probabilistic spatiotemporal model for reporting earthquake related events \cite{sakaki2013tweet}, developing a detection algorithm based on the frequency of tweets to detect earthquake \cite{earle2012twitter}, applying classifier methods on tweets to detect earthquake \cite{sakaki2010earthquake,robinson2013sensitive}, using qualitative approach to analyze people's behavior after an earthquake \cite{miyabe2012use}, applying a keyword level analysis to track social attitudes during and after an earthquake \cite{doan2011analysis}, and analyzing dynamic of rumor mill in tweets \cite{oh2010exploration}.

	Literature also contains methods for disaster management during hurricanes and typhoon. This research area analyzed online public communications by police and fire departments in Twitter and Facebook \cite{hughes2014online}, used classification methods to detect fake images during the hurricane \cite{gupta2013faking}, and analyzed external factors such as geolocation and internal factors (e.g. geolocation, and stakeholders power) in a small sample of tweets to find patterns of information dissemination \cite{takahashi2015communicating}. Recently, a study mapped online users' sentiments to track their changes during a hurricane \cite{neppalli2017sentiment}. Twitter data was also used to help disaster management using qualitative content analysis to categorize information inside the Twitter data during a volcano eruption \cite{sreenivasan2011tweet}.

	Flood-related studies developed numerous frameworks to improve disaster management during floods. These studies tracked users' behavior during a flood \cite{palen2010twitter}, combined  georeferenced social media messages and geographic features of flood phenomena \cite{de2015geographic}, used machine learning techniques to find tweets reporting damage  \cite{ashktorab2014tweedr}, mapped geo-tagged tweets to directly support geotechnical experts for reconnaissance purposes \cite{dashti2014supporting}, analyzed crisis relevant information from Twitter using related keywords and geo-tagged tweets \cite{herfort2014does}, investigated the correlation between location and the types of URLs inside tweets \cite{murthy2013twitter}, explored the discursive aspects of Twitter communication with qualitative analysis \cite{shaw2013sharing}, analyzed the related tweets with qualitative data coding to find useful information \cite{vieweg2010microblogging}, and used keyword analysis to determine the types of information in the flood-related tweets \cite{kongthon2012role}.

	Although it is evident from the literature that past research on the use of social media during a natural disaster provides useful insight into disaster management, unpredictable nature of natural disasters provides a great motivation for disaster responders to constantly improve disaster management by exploring new perspectives. Thus, we believe we can improve disaster management through better integration of SA into the management practices specially when it comes to people concerns and negative feelings. To our knowledge no other research studied the concerns of people behind their negative feelings for having a better situational awareness. This study proposes an analytical framework to explore a large number of tweets using sentiment analysis and topic modeling for tracking temporal patterns of public negative concerns during the 2015 SC flood.
	
	\section{Methodology and Results}
	\label{ME}

	This paper proposes a framework, \textit{\textbf{\underline{Twi}}tter \textbf{\underline{S}}ituational \textbf{\underline{A}}wareness (TwiSA)}, with four components: data collection, sentiment analysis, temporal topic discovery, and topic content analysis.

	\subsection{Data Collection}

	There are two methods to collect a large number of tweets in TwiSA. One method is to retrieve data using Twitter APIs (Application Programming Interfaces) \cite{twitapi}, and the other method is to ask an independent service to provide the data. The first method, despite being free, is limited in that one can only retrieve a small portion of relevant tweets. Therefore, we asked an independent company, Gnip \cite{gnip}, to extract 100\% of tweets related to our research. Similar to Palen et al. (2010)  \cite{palen2010twitter}, we ordered the queries in Table \ref{tab:query} to collect tweets for the days having flood danger or flood side effect. One million tweets were filtered based on 13 days in October 2015. We chose this time frame because it represents a period that starts with  the highest amount of rainfall (October 3, 2015) and ends with the last day for boil advisory (October 15, 2015)  \cite{zurich}. It is worth mentioning that we did not remove retweets and the tweets containing URLs from our dataset but the word \textit{``rt"} and URLs were removed from the collected tweets.
	Table \ref{tab:tweet_exp} shows a sample of tweets that are related to insurance and damage issues using the \#scflood query.

	\begin{table}[htp]
		
		\caption{Queries for Twitter Data Collection}
		\begin{center}
			\begin{tabular}{  p{14cm}  } 
				\hline
				
				\#floodsc \textbf{OR} \#scflood2015 \textbf{OR} \#SCFloodRelief \textbf{OR} \#southcarolinastrong \textbf{OR} \#prayforsc \textbf{OR} \#scflood \textbf{OR} \#scflooding \textbf{OR} \#FloodGSSCMMwithlove \textbf{OR} \#floodingsc \textbf{OR}  \#flood \textbf{OR} flood    \\  			\hline

			\end{tabular}
			
			\label{tab:query}
		\end{center} 
	\end{table}

	\begin{table}[htp]
		
		\caption{A Sample of Tweets}
		\begin{center}
			\begin{tabular}{  p{1.5cm}p{12cm}  } 
				\hline
				
				Tweet 1 & \textit{How many South Carolinians have flood insurance?  Few.They'll be looking to federal gov't or their tight-fisted governor for help. \#SCFlood} \\  	\hline
				Tweet 2	    &   \textit{Damage inside Student Activities room at Westwood. \#SCFlood}			\\ \hline

			\end{tabular}
			
			\label{tab:tweet_exp}
		\end{center} 
	\end{table}

		\subsection{Sentiment Analysis}

		Sentiment analysis discloses the overall feelings inside text data \cite{taboada2011lexicon}. Learning-based and lexicon-based methods are two main approaches for sentiment analysis \cite{khan2015combining}. The first approach uses machine learning classifiers when there is prior knowledge about data categories. In this case, a sample of the data is first labeled by human raters such as assigning spam and non-spam labels to a sample of emails \cite{taboada2011lexicon}. The second approach, a cost-effective one, finds the frequency of a predefined dictionary of positive and negative terms to disclose sentiment in the data when there is not any prior knowledge about its categories \cite{medhat2014sentiment}. We did not have any prior knowledge about the categories of the tweets in this research; therefore, we applied the second approach to find positive, negative, and neutral tweets.

		Linguistic Inquiry and Word Count (LIWC) \cite{liwcco} is a linguistics analysis tool that reveals thoughts, feelings, personality, and motivations in a corpus based on lexicon-based approach. This software assumes that the frequency of words can reveal overall tone in the corpus \cite{karami2014improving,karami2014exploiting,karami2015online}. This tool applies a simple word-count process in text documents and maps each word in the documents on the already developed internal LIWC dictionaries in different categories such as negative emotion \cite{franklin2015some}.
		
\begin{figure} [ht]%
			\centering
			\begin{tikzpicture}     
			\begin{axis}[    
			xbar, 
			height = 8cm,
			bar width=45pt,
			y axis line style = {opacity = 0 },
			y=-0.5cm,
			bar width=0.2cm,
			enlarge y limits=auto,
			ytick pos=left,
			nodes near coords, 
			nodes near coords align={horizontal},
			legend style={at={(0.5,1.1)},
				anchor=north,legend columns=-1},
			reverse legend,
			xticklabels={,,},
			extra y tick style={grid=none},
			width  = 12cm,
			hide x axis,
			ytick={10/15/2015,10/14/2015,10/13/2015,10/12/2015, 10/11/2015, 10/10/2015, 10/09/2015,10/08/2015, 10/07/2015, 10/06/2015, 10/05/2015, 10/04/2015, 10/03/2015},
			symbolic y coords = {10/03/2015, 10/04/2015,10/05/2015,10/06/2015,10/07/2015,10/08/2015,  10/09/2015,10/10/2015, 10/11/2015, 10/12/2015, 10/13/2015, 10/14/2015, 10/15/2015   },
			]
			\addplot [pattern=horizontal lines gray
			,pattern color=orange] coordinates {(16198 ,10/03/2015) (18710,10/04/2015)  (30022,10/05/2015)  (20319,10/06/2015)   (17803,10/07/2015)  (17575,10/08/2015) (15632,10/09/2015) (12745,10/10/2015) (12783,10/11/2015) (13208,10/12/2015) (13813,10/13/2015)  (14839,10/14/2015) (13427,10/15/2015)  };
			\end{axis}
			\end{tikzpicture}
			\caption{Number of Negative Tweets per Day}
			\label{fig:distribution}
		\end{figure}
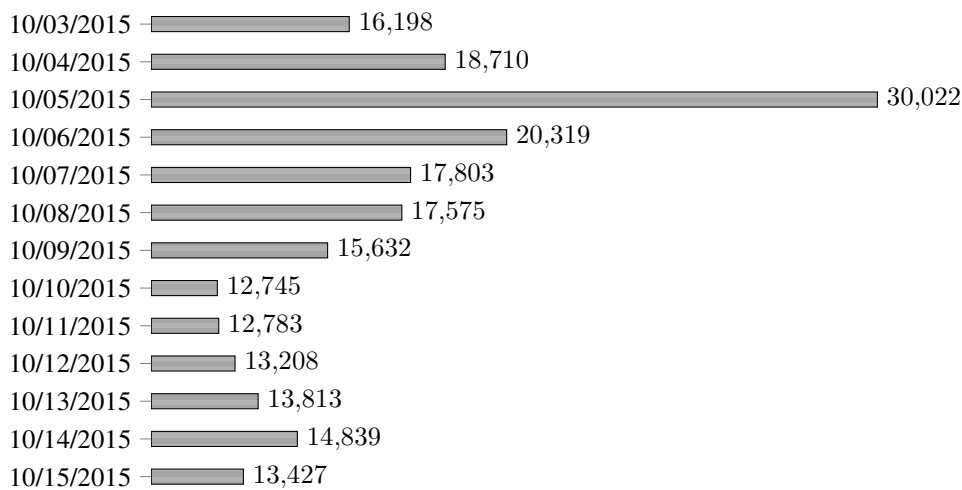

		LIWC has good sensitivity value, specificity value, and English proficiency measure \cite{golder2011diurnal}. Comparing to competitors such as deep learning, this tool can be used alone or be combined with other methods \cite{gonccalves2013comparing}. Applying LIWC on the collected tweets shows that there are 217,074 negative tweets, 529,150 neutral tweets, and 217,183 positive tweets. When people are exposed to unpredictable events, they show unpleasant negative feelings that come from the presence of threats with undesirable outcomes \cite{nolen1991prospective,yin2014anxious}. We used the negative tweets for the next step to identify unpleasant elements of the SC flood through temporal topic discovery analysis (Figure \ref{fig:sa}). Figure \ref{fig:distribution} shows that the number of negative tweets per day has changed from $\sim$12,700 tweets to $\sim$30,000 tweets with average $\sim$16,700 tweets per day.

		\begin{figure}[H]
			\centering
			\large
			\scalebox{0.6}{
				
				\begin{tikzpicture}[grow=down, -stealth]
				\node[bag]{Twitter} 
				child{ edge from parent node[]{}; \node[bag]{\textbf{\underline{23.18\% Negative Tweets}}}
				}
				child{ edge from parent node[]{}; \node[bag]{53.63\% Neutral Tweets}
				}
				child{ edge from parent node[]{}; \node[bag]{23.19\% Positive Tweets}
				};
				\end{tikzpicture}}
			\caption{Sentiment Analysis of the SC Flood Tweets}
			\label{fig:sa}
		\end{figure}
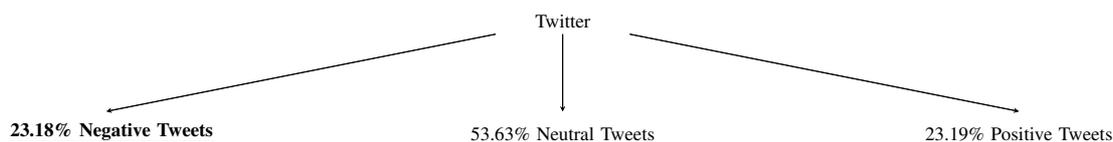

\subsection{Temporal Topic Discovery}
Next we turned our attention to detect topics among the pool of the \textbf{217,074} negative tweets to showcase the proposed framework. Different approaches for detecting the topics in a corpus have been developed based on the neural network and statistical distributions. During the last decade, the neural network has been considered for improving data analysis methods such as the deep neural network (DNN) \cite{zhang2016learning}. The models based on neural network are costly and time-consuming methods \cite{mohammadi2018deep,zhao2017research,chilimbi2014project} that require a lot of training data along with having a chance of overfitting problem \cite{erhan2010does}. For example, one study needed to train millions of tweets for the pre-training step \cite{severyn2015twitter} that is also a barrier for having an efficient real-time streaming data \cite{mahmud2018applications}. In addition, the deep learning methods haven't shown significant performance over other methods in small or medium size corpora \cite{mahmud2018applications,zhang2016learning}.

According to the literature, Latent Dirichlet Allocation (LDA) \cite{blei2003latent} is a valid \cite{mcauliffe2008supervised,hong2010empirical,karami2014fftm,karami2015fuzzyiconf} and widely used \cite{lu2011investigating,paul2011you} model for discovering topics in a corpus \cite{karami2018computational}. LDA assumes that there are multiple topics in a corpus and each topic is a distribution over corpus's vocabulary and can be inferred  from word-document co-occurrences \cite{blei2003latent,karami2015fuzzy,karami2018fuzzy}.

LDA assigns each word in a document to each of topics with a different degree of membership using a statistical distribution \cite{karami2015flatm}. For example, in a given corpus, LDA assigns ``flood," ``earthquake," and ``hurricane" into a topic with ``natural disaster" theme. TwiSA uses LDA to identify topics in the negative tweets. This research uses Mallet implementation of LDA \cite{mallet} that uses Gibbs sampling that is a specific form of Markov Chain Monte Carlo (MCMC) \cite{steyvers2007probabilistic} with removing standard  stopwords list and having 1000 iterations to detect top 25 topics per day (325 topics in total) in the negative tweets per each of the 13 days.

	\begin{table}[ht]
		\centering			
		
		\Large

		\caption{Flood Topics Examples}
		\scalebox{0.55}{
			\begin{tabular}{ cccccc }\hline

				\rowcolor[gray]{.9} \textbf{Victims} & \textbf{Damage \& Costs} & \textbf{Drinking Water} & \textbf{Insurance} & \textbf{Road Damage} & \textbf{Roof Damage}      \\\hline
				
				scflood &  damage        & water              & families        & road  & roof     \\
				
				victims & property    & drink            & insurance              & damage  & damage   \\
				
				flooding    & loss     & boil             & homeowners          & roadway   & repair  \\
				
				disaster   & roadway        & bottle            & destroyed        & st & danger    \\
				
				death       & construction   & clean           & support    & exit  & home  \\ 
				
				\hline	
				\rowcolor[gray]{.9} \textbf{Bridge Damage} & \textbf{Flood Report} & \textbf{Homelessness} &  \textbf{Power Lost}  &  \textbf{Animal} &      \\ \hline
				
				disaster &   alert           & flood              & lost        &  Animal     \\
				
				bridge   & warning    & homeless            & flood              & Shelter   \\
				
				flooded    & effect     & lives             & home          &  dog  \\
				
				natural   & flood        & woman            & power         &  poor    \\
				
				flood       & remain   & boy           & family    &  find  \\ 
				
				\hline					
				
			\end{tabular}}
			\label{tab:topicexamples}
		\end{table}
		
		\subsection{Topic Content Analysis}			
		
		The last component of TwiSA analyzes extracted topics. This component labels and categorizes 25 topics per day, analyzes frequency of them, and explains possible reasons behind the detected topics comparing with official and published reports. 
		
		We manually labeled the negative topics (concerns) for each of the 13 days between 10/03/2015 and 10/15/2015. For example, if \textit{``road"}, \textit{``damage"}, \textit{``roadway"}, \textit{``st"}, and \textit{``exit"} are in a topic, we labeled the topic \textit{``Road Damage"}.  Table \ref{tab:topicexamples} shows some examples of the detected topics. It is worth mentioning that the quantity of topics/day were more than 11 topics but we assigned related topics to one category. For example, if some topics represent injured people and victims, we assigned just ``Victims" label to each of those topics.

\begin{figure}[ht]
				\centering
				
				\begin{tikzpicture}[scale=0.7]
				\begin{axis}[ xlabel={Date}, ylabel={Number of Topics}, y label style= {at={(-0.1,0.5)}},legend style={at={(0.5,-0.2)}, anchor=north,legend columns=-1},
				ytick={1,2,3,4,5,6,7,8},
				symbolic y coords = {1,2,3,4,5,6,7,8},
				xtick={1,2,3,4,5,6,7,8,9,10,11,12,13},
				symbolic x coords = {1,2,3,4,5,6,7,8,9,10,11,12,13}]
				\addplot coordinates {(1,3)(2,5)(3,7) (4,8)(5,8) (6,8) (7,7)(8,5)(9,4)(10,3)(11,3)(12,3)(13,2)};
				
				\end{axis}
				\end{tikzpicture}
				\vspace{-3mm}
				\caption{Number of Negative Topics Per Day}
				\label{fig:numtopperday}
			\end{figure}
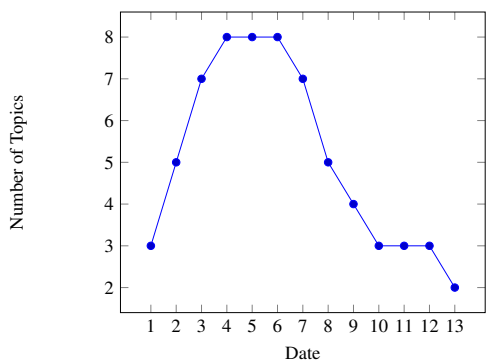

		The analyzed topics were categorized into 11 unique ones with different frequencies (Table \ref{tab:topfreq}). This analysis indicates that damages and animal support are the most and the least discussed negative topics, respectively.  Moreover, figure \ref{fig:numtopperday} shows that the number of unique negative topics per day has changed from 2 to 8 during the time period considered for this research. The highest number of topic diversity is on the fourth, the fifth, and the sixth days, and the lowest number of topic diversity is on the last day. This figure indicates growing rate of topic diversity between the first day and the sixth day, and declining rate of topic diversity between the seventh day and the last day. It seems that SC disaster responders have tried to reduce people's negative concerns between the sixth day and the thirteenth day.

	\begin{table}[ht]
		\centering
		\caption{Total  Frequency of Topics}
		\scalebox{0.95}{
			\begin{tabular}{|cc|cc| } \hline	
				{\cellcolor[gray]{.9}} \textbf{Topic} & {\cellcolor[gray]{.9}} \textbf{Percentage} & {\cellcolor[gray]{.9}} \textbf{Topic} &  {\cellcolor[gray]{.9}} \textbf{Percentage}   \\ 	\hline
				Victims & 18.75\% & Roof Damage &  9.37\% \\  \hline
				Damage and Costs  & 15.62\% & Bridge Damage  & 6.25\% \\  \hline
				Drinking Water  & 9.37\%& Flood Report   & 6.25\% \\  \hline
				Insurance  & 9.37\% & Power Lost  & 4.68\% \\  \hline
				Homelessness  & 9.37\% & Animal  &  1.6\% \\  \hline
				Road Damage  & 9.37\% &    &   \\  \hline

			\end{tabular}}
			
			\label{tab:topfreq}
		\end{table}

			Table \ref{tab:topics} shows the topics' distribution for each of the 13 days. We compare our results with official and published reports to explain possible reasons behind the detected topics.

		\begin{table}[ht]
			
			\caption{Negative Topics from 10/03/2015 to 10/15/2015}
			\begin{center}
				\scalebox{0.7}{
					\begin{tabular}{  |c|c|c|c|c|c|c|c|c|c|c|c|c|c|} 
						\hline
						
						& {\cellcolor[gray]{.9}} \rotatebox{90}{Day 1:  10/03/15} & 
						{\cellcolor[gray]{.9}} \rotatebox{90}{Day 2: 10/04/15}& {\cellcolor[gray]{.9}} \rotatebox{90}{Day 3: 10/05/15}& {\cellcolor[gray]{.9}} {\cellcolor[gray]{.9}} \rotatebox{90}{Day 4: 10/06/15}& 
						{\cellcolor[gray]{.9}} \rotatebox{90}{Day 5: 10/07/15}& {\cellcolor[gray]{.9}} \rotatebox{90}{Day 6: 10/08/15}& {\cellcolor[gray]{.9}} \rotatebox{90}{Day 7: 10/09/15}& {\cellcolor[gray]{.9}} \rotatebox{90}{Day 8: 10/10/15}& {\cellcolor[gray]{.9}} \rotatebox{90}{Day 9: 10/11/15}& {\cellcolor[gray]{.9}} \rotatebox{90}{Day 10: 10/12/15}& {\cellcolor[gray]{.9}} \rotatebox{90}{Day 11: 10/13/15}& {\cellcolor[gray]{.9}} \rotatebox{90}{Day 12: 10/14/15}& {\cellcolor[gray]{.9}} \rotatebox{90}{Day 13: 10/15/15} \\ \hline
						
						\textbf{Animal} & 0 & 1 &0 & 0&0 &0 &0 & 0&0 &0 &0 &0 &0  \\ \hline
						\textbf{Bridge Damage} & 0 &0 &1 & 0& 0&1 &1 &0 &0 & 0& 0&1 & 0 \\ \hline
						\textbf{Damage and Costs} & 1 &1 & 1 &1 &1 &0 &1 &1 &1 &0 &0 &1 & 1 \\ \hline
						\textbf{Drinking Water}& 0 & 0&1 &1 & 1& 1&1 &1 & 0& 0&0 & 0& 0 \\ \hline
						\textbf{Flood Report} & 1 & 1& 0&1 &1 &0 & 0&0 &0 &0 &1 &0 &0  \\ \hline
						\textbf{Homelessness} & 1 & 0& 1&1 &1 & 1& 1&0 & 0& 0&0 &0 & 0 \\ \hline
						\textbf{Insurance} & 0 &0 & 0&1 &1 &1 &0 &0 &1 &1 &1 &0 &0  \\ \hline
						\textbf{Power Loss} & 0 & 1&0 &0 &0 &1 &0 &1 &0 &0 &0 &0 &0  \\ \hline
						\textbf{Road Damage} &0  &0 &1 &1 &1 &1 &1 &0 &0 &1 &0 &0 &0  \\ \hline
						\textbf{Roof Damage} & 0 & 0& 1&1 &1 &1 &1 &1 &0 &0 &0 &0 &0  \\ \hline
						\textbf{Victims} & 0 & 1&1 &1 &1 &1 &1 &1 &1 &1 &1 &1 &1  \\ \hline

					\end{tabular}}
					
					\label{tab:topics}
				\end{center} 
			\end{table}

			``Victims" is the most frequent negative topic that is about the people who were harmed, injured, or killed during the SC flood. South Carolina Emergency Management Department (SCEMD) develops and coordinates the South Carlina emergency management program for effective response to disasters. SCEMD reported that 19 people were killed and more than 1,500 were rescued from water \cite{scemd}. 
			
			``Damage \& Costs" is the next high frequent topic that describes financial damage. SCEMD reported that 28,162 people received \$2.2 billion aid through FEMA (Federal Emergency Management Agency) and SCEMD. In addition, more than 3000 collisions \cite{scemd} and more than 73,000 damaged structures \cite{zurich} were recorded during the SC flood.

			The next topic is ``Drinking Water". During the 2015 SC flood, the drinking water system was collapsed in some area such as Columbia (South Carolina's capital city) \cite{pjstar} and thousands of SC residents did not have water for days \cite{state1}.  In addition, boil water advisory was issued for some days in some parts of South Carolina \cite{cola}.  With respect to ``Insurance" topic, most SC residents did not have flood insurance in 2015 and FEMA allocated more than \$89 million for individual assistance \cite{prnews}. In addition, it was suggested to increase insurance penetration and accessibility in South Carolina \cite{zurich}. 
			
			The next topic is ``Homelessness." Thousands of SC residents were displaced and hundreds of people were staying in shelters \cite{state1}. This topic covers two groups of people. First people who were forced to leave their home and the second people who did not have a permanent place to live \cite{wbtv}. This topic is an interesting one because SCEMD report did not mention this topic \cite{scemd}.

			``Road", ``Roof", and ``Bridge" damages are the next three issues. South Carolina has the fourth largest state-owned highway system with 41,000 miles of road and 8,400 bridges in U.S.A \cite{statelib}. More than 500 roads and bridges were damaged during the flood \cite{scemd}, and 8 out of the 19 deaths occurred on a flooded road \cite{weather}. The SC flood caused expensive roof damage. The estimates placed the repair cost close to \$137 million dollars \cite{greenville,postandcourier,zurich}. Hundreds of bridges were damaged and closed in October 2015 \cite{scemd}. 
			
			``Flood Report" topic shows the reports and warnings that were retweeted (rt). 41 warning were issued during the flood, and different contents such as images from the floods, local storm reports, and areas of major flooding were shared \cite{weather}. The next topic is ``Power Loss". Thousands of people did not have power for some days during the flood \cite{state1}. Finally, ``Animals" was the last and the least discussed topic in the negative tweets. This topic shows main public concerns for animals. During the 2015 SC flood, hundreds of pets were rescued \cite{cnbc} and some of them lost their homes \cite{masslive}. Supporting animals was one of the topics mentioned in tweets on the second day. This is the second interesting topic that was not addressed in the SCEMD's report. On the other side, we saw that food was the only topic that was in the SCEMD's report but not in our analysis. More than 2 million meals were served \cite{scemd} and it seems that this topic wasn't among main concerns during the flood.
			
			The findings show that this research helps situational awareness in some directions. The first one is providing a big picture of damages and public concerns. For example, detecting damage reports by our framework shows show the large size of damages because LDA identifies topics discussed by a large number of users. That big picture can aid the state and federal agencies to have a better-cost estimation and resource allocation. This study can give the agencies a list of priorities for each day. For example, we think that the considering and addressing the insurance concern in the first five days were an excellent strategy to calm the people without an immediate cost.
			
			Comparing to other studies, this study has several benefits. First,  TwiSA is a fast, real-time, and cost effective framework. The second benefit is combining sentiment analysis and topic modeling methods. While other studies have used the two methods separately without a connection between them in SA analysis, this research provides a flexible framework that has a potential to embed other techniques such as utilizing deep learning for sentiment analysis and LDA. The next benefit is focusing on disagreeable or negative experiences, instead of all experiences. The third advantage is introducing new issues such as insurance in a disaster, which were not considered in other studies and confirmed by official reports. The next asset is to have a dynamic perspective exploring the disaster issues during a time frame.

				Accessing high quality and  relevant real-time data during a natural disaster can increase safety and reduce damage and social effects \cite{kitchin2014real}. Every user is a sensor and contributor in social media and can generate valuable and immediate real-time data to develop better situational awareness. However, social media users generate a huge amount of data that needs to be summarized to provide a big picture in a disastrous situation. 	
				
				The opinion survey is a traditional method to analyze public opinion; however, this expensive method should be implemented after natural disasters. In addition, a low number of people participate in the data collection and data analysis steps takes a considerable amount of time \cite{urcan2012flood}. 
				
				The growth of social media has provided a great opportunity to track public opinion. Big real-time social media data can help disaster managers to develop a better SA during a natural disaster. This research proposes an analytical framework to detect and track leading public concerns on social networks such as Twitter to provide a better picture of a disaster. TwiSA combines sentiment analysis and topic modeling to handle a hug number of tweets and to disclose negative public concerns. 
				
				This paper found that ``victims", ``damages", ``drinking water", ``insurance", ``flood report", ``power loss", `homelessness", and ``animals" are the topics that were discussed in negative tweets during the flood. We also found two topics, ``homelessness", and ``animals", were not mentioned in the official reports after the flood. It seems that human and non-human damages, drinking water, and insurance  were the most discussed topics during the SC flood.  Even though TwiSA was applied to the 2015 SC flood twitter data, it is not limited to this context and  can be applied to other natural disasters and emergency situations.

				This research shows that social media are valuable data sources to explore people concerns during a natural disaster. TwiSA can help disaster management teams to find public's concerns in the fastest way to develop a better real-time crisis management plan. The proposed framework can be used not only during natural disasters but also after natural disasters as a post-event methodology review capability (PERC) for future disaster risk reduction. This research has applications for policymakers and disaster risk management experts in developing a better strategy for analyzing social media contents and responding to it. It also helps social scientific research in developing Public Service Announcement (PSA) and unveiling public opinions \cite{weather}.

				Our future directions center on minimizing the impact of noises and misinformation, and analyzing positive and neutral tweets. A potential limitation of our approach is that LIWC tool doesn't usually detect emotions, misspellings, colloquialisms, foreign words, sarcasm, and abbreviations \cite{pennebaker2015development}. Another limitation is related to the tweets containing wrong or inaccurate information and spams \cite{najafabadi2017research}. LDA detects major topics, not every single of topics, in a corpus. Therefore, we assumed that the massive number of tweets in a disaster buries noises. While we believe that the noises don't have a significant impact on the results, we will utilize some approaches to reduce the possible effects of the noises. The first approach is removing retweets and the tweets containing URL to avoid spams. The second approach is detecting unrelated or inaccurate topics and then eliminating the tweets whose primary topic is among the irrelevant or inaccurate topics. The primary topic is the topic that has the highest probability for a tweet. The next approach is using bot detection tools such as Botometer \cite{ferrara2016rise} to remove social bots spreading misinformation. We believe that the future plan will address these limitations to provide higher quality data and new insights to SA.

				\begin{acks}
					This research is supported in part by the University of South Carolina Office of the Vice President for Research through the 2015 SC Floods Research Initiative. The authors are grateful for this support and all opinions, findings, conclusions and recommendations in this article are those of the authors and do not necessarily reflect the views of the funding agency.
				\end{acks}

			\end{document}